# HiPPO: Exploring A Novel Hierarchical Pronunciation Assessment Approach for Spoken Languages

Bi-Cheng Yan[1], Hsin-Wei Wang[1], Fu-An Chao[1], Tien-Hong Lo[1],
Yung-Chang Hsu[2], Berlin Chen[1]

[1]National Taiwan Normal University, [2]EZAI
{bicheng, berlin}@ntnu.edu.tw

## Abstract

Automatic pronunciation assessment (APA) seeks to quantify a second language (L2) learner's pronunciation proficiency in a target language by offering timely and fine-grained diagnostic feedback. Most existing efforts on APA have predominantly concentrated on highly constrained reading-aloud tasks (where learners are prompted to read a reference text aloud); however, assessing pronunciation quality in unscripted speech (or free-speaking scenarios) remains relatively underexplored. In light of this, we first propose HiPPO, a h̲ierarchical p̲ronunciation assessment model tailored for s̲poken languages, which evaluates an L2 learner's oral proficiency at multiple linguistic levels based solely on the speech uttered by the learner. To improve the overall accuracy of assessment, a contrastive ordinal regularizer and a curriculum learning strategy are introduced for model training. The former aims to generate score-discriminative features by exploiting the ordinal nature of regression targets, while the latter gradually ramps up the training complexity to facilitate the assessment task that takes unscripted speech as input. Experiments conducted on the Speechocean762 benchmark dataset validates the feasibility and superiority of our method in relation to several cutting-edge baselines.

## 1 Introduction

Spurred by the global demand for foreign language proficiency in both the workforce and academia, computer-assisted pronunciation training (CAPT) has gained significant attention, which facilitates second-language (L2) learners to practice pronunciation skills with near-instant, instructive, and potentially diagnostic feedback (Norris and

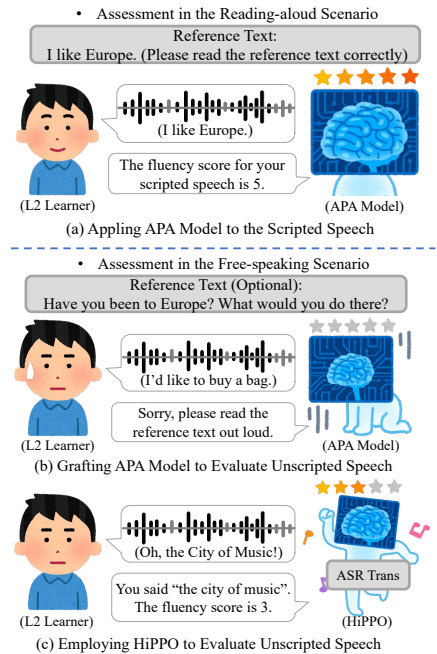

Figure 1: Outlines our motivations. (a) Existing APA models are primarily tailored for read-aloud tasks. (b) Directly applying APA models to free-speaking scenarios struggles to quantify oral skills based on speech signals. (c) HiPPO integrates a speech recognizer to generate transcriptions from the learner's speech, effectively reformulating free-speaking assessment as a task akin to read-aloud.

Davis, 2025; Moere and Downey, 2016). To meet this pressing demand, CAPT systems have become ubiquitous and appealing learning tools, transitioning the conventional pedagogical approach from teacher-led instruction to self-directed learning (Rogerson-Revell, 2021; Chen and Li, 2016; Singla et al., 2021).

Automatic pronunciation assessment (APA) aims to evaluate L2 learners' speaking proficiency and provide fine-grained feedback on specific pronunciation aspects pertaining to a target language, figuring prominently in the field of CAPT. Prior studies on APA have primarily drawn attention to highly constrained speaking tasks (such as listening and then repeating words or sentences). As exemplified in Figure 1(a), a defacto archetype system for APA is instantiated in reading-aloud (or scripted) learning scenarios, where an L2 learner is provided with a reference text and instructed to pronounce it correctly. Methods in this line of research typically rely on an input reference text paired with the learner's speech to derive timestamps of linguistic units (i.e., phones or words) via an automatic speech recognition (ASR) system, which are then used for either pronunciation feature extraction (Gong et al., 2022; Chao et al., 2022; Do et al., 2023; Yan et al., 2024) or for neural modeling (Lin and Wang, 2021; Wang et al., 2025). Albeit achieving competitive performance in relation to inter-rater agreement (Yan and Chen, 2024; Pei et al., 2024), scripted-speech assessments fail to reflect learners' speaking abilities in real-world communication. In contrast, pronunciation assessment of spoken languages introduces new challenges to CAPT, as it attempts to quantify an L2 learner's oral skills in spontaneous speech or elicit authentic responses through short questions (Zechner and Evanini, 2019; Kheir et al., 2023). Directly grafting existing APA models to use cases of spoken language assessment, however, confronts at least two major issues. First, as shown in Figure 1(b), the utterances of an L2 learner are produced in an unscripted manner, which makes APA models struggle to extract correct pronunciation features encompassing time-alignment information (Shen et al., 2021; Deng et al., 2020; Witt and Young, 2000). What is more, owing to the free-form nature of unscripted speech, the desired APA models are required to accommodate speech input of varying lengths.

Building on these observations, this paper presents HiPPO, a novel <u>hi</u>erarchical <u>p</u>ronunciation assessment model for s<u>po</u>ken languages that evaluates L2 learners' oral proficiency based on unscripted speech (or free-speaking scenarios) and provides analytical scores on various pronunciation aspects across multi-granular linguistic levels. Specifically, HiPPO strategically employs a speech foundation model along with a grapheme-to-phoneme (G2P) converter to derive the most likely phone sequence produced by an L2 learner, thereby bringing the assessment task closer to its scripted-speech counterpart, as illustrated in Figure 1(c). To overcome sequence length constraints and preserve articulatory traits across multi-granular linguistic units, HiPPO capitalizes on a tailor-made Conv-LLaMA block to stack a hierarchical neural architecture, which augments the LLaMA block (Touvron et al., 2023) with a convolutional branch and rotary position encoding (Su et al., 2024). Moreover, during training, a contrastive ordinal regularizer is put forward to modulate feature distances through the absolute differences between regression targets. By exploiting the ordinal constraints, the proposed regularizer serves as a promising approach to generate score-discriminative features, mitigating the detrimental effects of ASR errors on pronunciation assessments. We further introduce a simple yet effective curriculum learning strategy for HiPPO that progressively increases the training complexity, transforming the assessment tasks from the read-aloud scenario to the free-speaking counterpart. An extensive set of experiments conducted on Speechocean762 benchmark dataset (Zhang et al., 2021), consisting of both read-aloud and simulated free-speaking scenarios, demonstrates substantial and consistent performance gains of the proposed methods over several strong baselines.

In summary, our contributions are at least four-fold: (1) to our knowledge, HiPPO is the first attempt to assess oral skills for unscripted speech with multi-faceted scores from phone to utterance levels, opening a new avenue for CAPT; (2) we propose a novel Transformer block, Conv-LLaMA block, as the backbone of HiPPO, elaborately designed to handle the free-from speech uttered by L2 learners; (3) to alleviate the negative effects of ASR errors, a contrastive ordinal regularizer is proposed to reflect the ordinality of regression targets within the feature space; and (4) a simple yet effective curriculum learning strategy is explored to boost the performance of pronunciation assessment in the free-speaking scenario.

## 2 Methodology

This section sets out with a problem definition for pronunciation assessments on unscripted speech (or free-speaking scenarios) and then sheds light on the proposed methods, encompassing the assessment model, training objectives, and learning strategy. Due to the space limit, the overview of related work will be given in Appendix A.

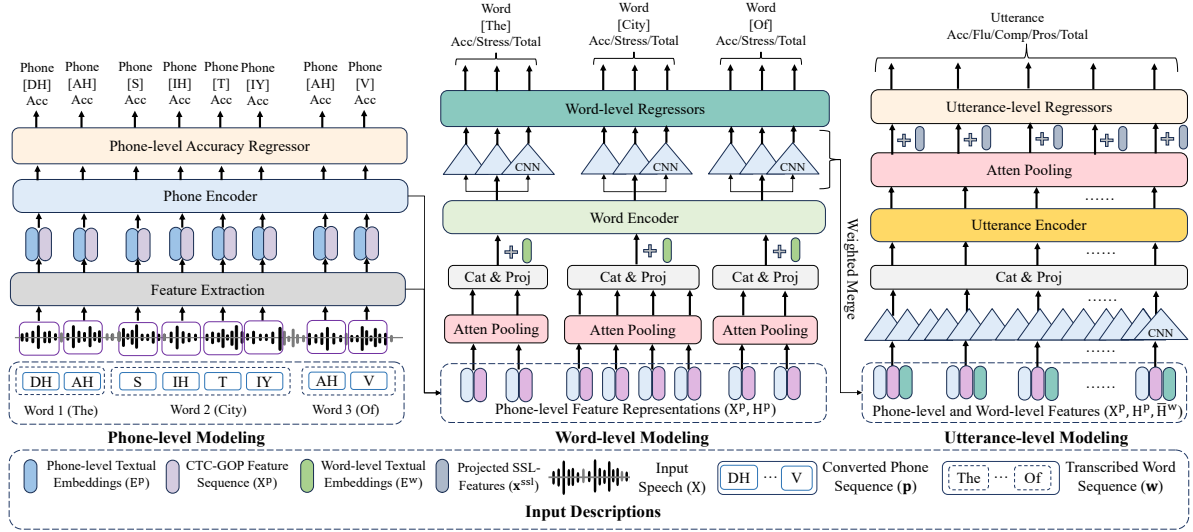

Figure 3: The overall architecture of the proposed hierarchical pronunciation assessment model (HiPPO).

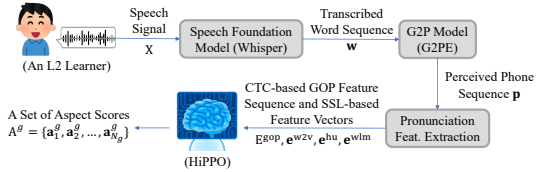

Figure 2: Processing flow of HiPPO for qualifying the oral skills in unscripted speech.

## 2.1 Problem Definition

To assess speaking skills across different linguistic granularities for unscripted speech, as illustrated in Figure 2, we first employ a speech foundation model[1] to transcribe a speech signal X produced by an L2 learner into a sequence of $M$ words $\mathbf{w} = (w_1, w_2, ..., w_M)$ and subsequently a G2P converter[2] to generate the corresponding phonetic transcription of $N$ phones $\mathbf{p} = (p_1, p_2, ..., p_N)$, where $\mathbf{w}$ and $\mathbf{p}$ collectively serve as a proxy for the textual and phonetic realizations perceived by human raters. Let $G = \{g^{phn}, g^{word}, g^{utt}\}$ denotes the set of linguistic granularities, where $g^{phn}$, $g^{word}$ and $g^{utt}$ mark the phone-, word-, and utterance-level linguistic granularities, respectively. HiPPO is trained under a multi-task learning paradigm to estimate a set of aspect score sequences $A^g = \{\mathbf{a}_1^g, \mathbf{a}_2^g, ..., \mathbf{a}_{N_g}^g\}$ for each granularity $g \in G$, where $N_g$ is the number of pronunciation aspects.

## 2.2 Hierarchical Pronunciation Assessment Model for Spoken Languages (HiPPO)

Figure 3 depicts the model architecture of HiPPO, which encompasses three major modeling stages: phone-, word-, and utterance-level modeling. In each of these modeling stages, the corresponding encoder is constructed with a newly proposed Conv-LLaMA block. After obtaining the representations of all pronunciation aspects, a distinct regressor is used to generate the pronunciation score of each aspect.

**Pronunciation Feature Extraction.** To portray the pronunciation quality of X, we extract connectionist temporal classification (CTC)-based goodness pronunciation (GOP) features for each phone in $\mathbf{p}$, where the pronunciation quality is measured as the likelihood ratio of all valid CTC alignments of $\mathbf{p}$ to that of the deviated phonetic transcripts (Cao et al., 2024). Compared to previous studies on the GOP feature extraction (Witt and Young, 2000; Hu et al., 2015; Shen et al., 2021), the CTC-based method computes GOP scores without explicit timestamps of phone segments and inherently tackles alignment errors by accounting for insertions and/or deletions in the deviated phonetic transcriptions. Additionally, to capture supra-segmental articulation cues and mitigate the data-sparsity issue frequently occurring in L2 speech corpora (Lo et al., 2024; Bannò and Matassoni, 2022), we leverage self-supervised learning (SSL)-based features for utterance-level pronunciation modeling. The SSL-based features are extracted at the frame-level and then aggregated to the utterance-level via simple mean pooling over time (Chao et al., 2022; Kim et al., 2022). A bit of terminology: the pronunciation feature extraction of HiPPO produces a phone-level pronunciation feature sequence $X^p \in \mathbb{R}^{d_p \times N}$ and a projected SSL-based feature vector $\mathbf{x}^{ssl} \in \mathbb{R}^{d_u \times 1}$, where $N$ is the length of the phone sequence, and $d_p$ and $d_u$ represent the hidden dimension of phone- and utterance-level modeling. The processing flow summarized as follows:

$$X^p = \text{Lin}_p(E^{gop}), \quad (1)$$

$$\mathbf{x}^{ssl} = \text{Lin}_{ssl}([\mathbf{e}^{w2v}; \mathbf{e}^{hu}; \mathbf{e}^{wlm}])), \quad (2)$$

where $\text{Lin}_p(\cdot)$ and $\text{Lin}_{ssl}(\cdot)$ are linear projections, and $[;]$ is a concatenation operation. $E^{gop} \in \mathbb{R}^{41 \times N}$ refers to the CTC-based GOP features extracted from a well-trained CTC-based ASR model [3], while $\mathbf{e}^{w2v}, \mathbf{e}^{hu},$ and $\mathbf{e}^{wlm} \in \mathbb{R}^{1024 \times 1}$ are utterance-level SSL-feature vectors derived from pre-trained acoustic models, viz. wav2vec-2.0, Hubert, and WavLM, respectively.

---

[1] https://huggingface.co/openai/whisper-large-v3

[2] https://github.com/Kyubyong/g2p

**Convolution-augmented LLaMA Block (Conv-LLaMA).** To model a pronunciation feature sequence of arbitrary length and capture nuanced articulation traits across linguistic units, we introduce a Conv-LLaMA block to stack a hierarchical assessment model, which enhances the model component of LLaMA (Touvron et al., 2023) with a convolutional branch and rotary position encoding. As depicted in Figure 4, the proposed block comprises two branches: one branch captures supra-segmental articulation cues via a multi-head self-attention (MHSA) module followed by a swish-gated linear unit (SwiGLU) operation (Touvron et al., 2023), while the other focuses on capturing local pronunciation traits via a convolutional neural network (CNN) module. Subsequently, these two branches are combined via a weighted average operation (Peng et al., 2022). The proposed CNN module is equipped with two key components, i.e., a point-wise convolution for capturing information across feature dimensions and a depth-wise convolution layer for extracting local spatial patterns. On the other hand, the MHSA module incorporates rotary position encoding (RoPE), a relative position encoding method developed for extrapolating feature sequence lengths, which operates through channel-wise multiplication on the key and query vectors in the multi-head self-attention layer (Su et al., 2024).

---

[3] https://github.com/frank613/CTC-based-GOP.git

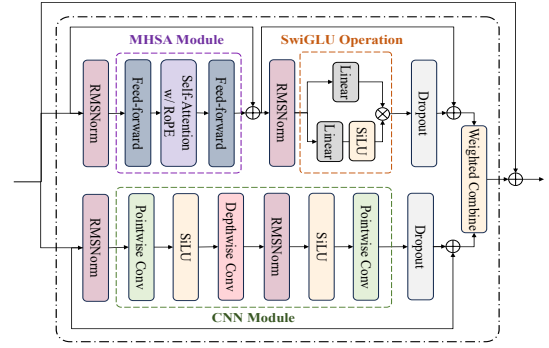

Figure 4: A schematic illustration of the proposed Conv-Llama block.

**Hierarchical APA Modeling.** For the phone-level assessment, we first combine the pronunciation features $X^p$ with the textual embeddings $E^p \in \mathbb{R}^{d_p \times N}$ in a point-wise manner, followed by a phone encoder to obtain aspect representations $H^p$:

$$H_0^p = X^p + E^p, \quad (3)$$

$$H^p = \text{PhnEnc}(H_0^p), \quad (4)$$

where $E^p$ is generated by passing phonetic transcription $\mathbf{p}$ into a phone embedding layer, and $\text{PhnEnc}(\cdot)$ is a stack of 3 Conv-LLaMA blocks. Subsequently, a regressor is built on top of $H^p$ to produce phone-level accuracy scores.

For word-level assessments, we begin by deriving a word representation vector from its constituent phones with a dedicated attention pooling, implemented with a 1-D depth-wise convolution layer, an MHA layer, and an average operation. The word-level input features $X^w \in \mathbb{R}^{d_w \times M}$ are obtained by feeding $X^p$ and $H^p$ through word-level attention pooling, and then packing their pooled counterparts together via a linear projection:

$$\hat{X}^w = \text{AttPool}_{w_1}(X^p), \quad (5)$$

$$\hat{H}^w = \text{AttPool}_{w_2}(H^p), \quad (6)$$

$$X^w = \text{Lin}_w([\hat{X}^w; \hat{H}^w]), \quad (7)$$

where $M$ denotes the length of transcribed word sequence, and $d_w$ symbolizes the hidden dimension of word-level modeling[4]. Following the integration of word-level textual embeddings $E^w$ with $X^w$, a word encoder is employed to generate a

---

[4] For efficient parallel computation, a word-level representation is duplicated to length of constituent phones.

sequence of contextualized representations $H^w \in \mathbb{R}^{d_w \times M}$:

$$H_0^w = X^w + E^w, \quad (8)$$

$$H^w = \text{WordEnc}(H_0^w), \quad (9)$$

where $E^w$ are obtained by mapping the transcribed word sequence **w** through modernBERT (Warner et al., 2024), and WordEnc(·) consists of 2 Conv-LLaMA blocks. Consequently, three distinct 1-D depth-wise convolution layers are performed on top of $H^w$ to generate aspect representations (viz. $H^{w_1}$, $H^{w_2}$, and $H^{w_3}$). The word-level pronunciation scores (accuracy, stress, and total) are generated by passing the aspect representations into the corresponding regressors.

For the utterance-level assessments, we first fuse $H^{w_1}$, $H^{w_2}$, and $H^{w_3}$ with a weighted average operation to produce $\bar{H}^w \in \mathbb{R}^{d_w \times M}$. After the distinct forward propagation through 1-D depth-wise convolution layers on $X^p$, $H^p$, and $\bar{H}^w$, the corresponding outputs are combined via a linear projection, and then fed into an utterance encoder to generate contextualized representations $H^u$:

$$\bar{H}^w = \text{Merge}(H^{w_1}, H^{w_2}, H^{w_3}), \quad (10)$$

$$H_0^u = \text{Lin}_u([\text{DC}_1(X^p); \text{DC}_2(H^p); \text{DC}_3(\bar{H}^w)]), \quad (11)$$

$$H^u = \text{UttEnc}(H_0^u), \quad (12)$$

where UttEnc(·) is a single Conv-LLaMA block, and $\text{DC}_1(\cdot)$, $\text{DC}_2(\cdot)$, and $\text{DC}_3(\cdot)$ are distinct 1-D depth-wise convolution layers, each of which has a kernel size of 3. Afterward, five separate attention pooling layers are stacked on top of $H^u$ and then integrated with the projected SSL-based feature vector $x^{ssl}$ via separate residual connections. These aspect representation vectors are processed by the corresponding regressors to derive the utterance-level aspect scores (viz. accuracy, fluency, completeness, prosody, and total).

### 2.3 Training Objectives

For the proposed model, we first consider a weighted sum of mean squared error (MSE) losses as the training objective, collected from multiple aspects across granularities:

$$\mathcal{L}_{APA} = \sum_{g \in G} \lambda_g \times \frac{1}{N_g} \sum_{k=0}^{N_g-1} \mathcal{L}_{g_k}, \quad (13)$$

where $\lambda_g$ denotes adjustable parameter, $N_g$ is number of aspects at granularity $g$, and $\mathcal{L}_{g_k}$ represents the MSE loss computed for the $k$-th aspect score sequence.

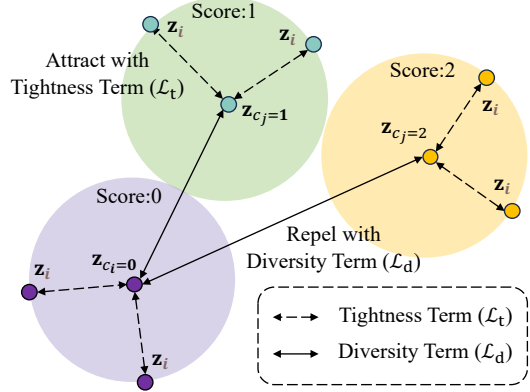

Figure 5: Illustration of contrastive ordinal (CONO) regularizer, which preserves inter-score discrepancies with diversity term $\mathcal{L}_d$ and maintains intra-score compactness via the tightness term $\mathcal{L}_t$.

**Constative Ordinal Regularizer.** To mitigate the detrimental effects of ASR errors on assessment performance, we devise a contrastive ordinal (CONO) regularizer to extract score-discriminative features. As phone-level representations are essential for constructing a hierarchical assessment model, we first extract an utterance-level feature **z** by averaging the outputs of the phone-level encoder $H^p$ over time. For a training batch of $L$ utterances, the corresponding feature vectors are aggregated to form a sequence $Z = (z_1, z_2, \ldots, z_L)$.

As depicted in Figure 5, the CONO regularizer encourages the feature vectors Z to render the ordinal relationship of the utterance-level accuracy scores $\mathbf{y} = (y_1, y_2, \ldots, y_L)$ via the synergy of a diversity term $\mathcal{L}_d$ and a tightness term $\mathcal{L}_t$:

$$\mathcal{L}_{CONO} = \lambda_d \mathcal{L}_d + \lambda_t \mathcal{L}_t, \quad (14)$$

where $\lambda_d$ and $\lambda_t$ are trade-off parameters. The diversity term $\mathcal{L}_d$ preserves inter-score discrepancies by minimizing the negative distances between score centroid vectors $z_{c_i}$ with a penalty:

$$\mathcal{L}_d = -\frac{1}{M(M-1)} \sum_{i=1}^{K} \sum_{i \neq j} w_{ij} \left\| z_{c_i} - z_{c_j} \right\|_2, \quad (15)$$

where $K$ is the number of score centers, and penalty $w_{ij} = |y_i - y_j|$ signifies the absolute differences between the regression targets. The score centroid vectors $z_{c_i}$ and $z_{c_j}$ are computed from Z by averaging all feature vectors whose utterance-level accuracy scores are $y_i$ and $y_j$, respectively. The tightness term $\mathcal{L}_t$ regulates intra-score compactness by pulling feature

| Models | Phone Scores | | Word Score (PCC) | | Utterance Score (PCC) | | | |
|---|---|---|---|---|---|---|---|---|
| | MSE↓ | PCC↑ | Accuracy↑ | Total↑ | Accuracy↑ | Fluency↑ | Prosody↑ | Total↑ |
| Liu2023 | - | - | - | - | - | 0.795 | - | - |
| VanillaSSL | - | - | - | - | 0.692 (±0.006) | 0.757 (±0.010) | 0.757 (±0.009) | 0.714 (±0.006) |
| MultiPA | - | - | 0.427 (±0.008) | 0.436 (±0.010) | 0.705 (±0.009) | 0.772 (±0.010) | 0.763 (±0.016) | 0.730 (±0.006) |
| Parallel-TFR | 0.240 (±0.003) | 0.330 (±0.009) | 0.416 (±0.016) | 0.417 (±0.019) | 0.717 (±0.014) | 0.797 (±0.003) | 0.791 (±0.003) | 0.741 (±0.010) |
| Parallel-LLaMA | 0.237 (±0.001) | 0.345 (±0.004) | 0.426 (±0.012) | 0.428 (±0.011) | 0.726 (±0.006) | 0.799 (±0.006) | 0.791 (±0.005) | 0.748 (±0.004) |
| Hier-LLaMA | 0.238 (±0.001) | 0.328 (±0.008) | 0.412 (±0.011) | 0.418 (±0.012) | 0.692 (±0.012) | 0.786 (±0.008) | 0.780 (±0.006) | 0.724 (±0.008) |
| HiPPO | **0.202** (±0.003) | **0.480** (±0.013) | **0.520** (±0.016) | **0.521** (±0.016) | **0.733** (±0.006) | **0.806** (±0.003) | **0.797** (±0.002) | **0.754** (±0.006) |
| w/o CONO | 0.213 (±0.004) | 0.448 (±0.012) | 0.513 (±0.007) | 0.516 (±0.007) | 0.720 (±0.005) | 0.797 (±0.003) | 0.791 (±0.002) | 0.743 (±0.005) |
| w/o CL | 0.241 (±0.002) | 0.331 (±0.011) | 0.404 (±0.012) | 0.404 (±0.014) | 0.698 (±0.010) | 0.790 (±0.011) | 0.785 (±0.011) | 0.728 (±0.007) |

Table 1: The performance evaluations of our model and all compared methods on Speechocean762 test set in simulated free-speaking scenarios.

representations $\mathbf{z}_i$ towards their score centroid vectors $\mathbf{z}_{c_i}$:

$$\mathcal{L}_t = \frac{1}{L}\sum_{i=1}^{L}\|\mathbf{z}_i - \mathbf{z}_{c_i}\|_2. \quad (16)$$

The training objective of HiPPO is designed as a linear combination of the pronunciation assessment task $\mathcal{L}_{APA}$ and the CONO regularization $\mathcal{L}_{CONO}$:

$$\mathcal{L} = \mathcal{L}_{APA} + \lambda_{CONO}\mathcal{L}_{CONO}, \quad (17)$$

where $\lambda_{CONO}$ is a tunable hyperparameter.

### 2.4 Curriculum Learning

Drawing inspiration from education systems, curriculum learning techniques improve model performance by progressively escalating training complexity from simple to hard (Bengio et al., 2009; Castells et al., 2020; Vakil and Amiri, 2023). The proposed curriculum training strategy starts from assessing pronunciation in a reading-aloud scenario $\mathcal{L}_{read}$, and gradually shifts towards assessing pronunciation in the free-speaking counterpart $\mathcal{L}_{free}$. In $\mathcal{L}_{read}$, the pronunciation features are extracted from the learner's speech alongside the corresponding reference text, while in $\mathcal{L}_{free}$ the transcribed word sequence serve as an alternative for pronunciation feature extraction. At each training iteration $\tau$, HiPPO selects a task from $\mathcal{L}_{read}$ with a probability of $1 - \mathcal{P}(\tau)$, or from the $\mathcal{L}_{free}$ with a probability of $\mathcal{P}(\tau)$, where $\mathcal{P}(\tau) = \tau/T$ is a scheduling function, with $T$ being the total number of training iterations and $\tau \in [0, T]$. The training strategy at iteration $\tau$ is defined by

$$(1 - \mathbb{I}(\tau))\mathcal{L}_{read} + \mathbb{I}(\tau)\mathcal{L}_{free}, \quad (18)$$

with the indicator function $\mathbb{I}(\tau)$ given by

$$\mathbb{I}(\tau) = \begin{cases} 1, \text{learning hard task } (w.p.\mathcal{P}(\tau)) \\ 0, \text{learning easy task } (w.p. 1 - \mathcal{P}(\tau)) \end{cases}. \quad (19)$$

## 3 Experimental Settings

This section describes the benchmark dataset and metrics used in this paper. Implementation details and descriptions of comparative methods are elaborated in Appendices B and C. Furthermore, HiPPO and the experimental dataset are publicly available to ensure the reproducibility of our work, accelerate CAPT research, and facilitate standardized evaluation[5].

**Benchmark Dataset.** A series of experiments were carried out on the Speechocean762 dataset, a publicly available corpus specifically designed for CAPT research (Zhang et al., 2021). This dataset comprises 5,000 English-speaking recordings collected from 250 Mandarin L2 learners, with training and test sets of equal size, each containing 2,500 utterances. Speechocean762 was collected in a reading-aloud scenario (reading reference texts aloud) with accessible reference texts and corresponding canonical phones (phone-level reference text). To simulate a free-speaking scenario for possible use cases of spoken language assessment, we exclude these reference texts from the model input and rely instead on the ASR transcribed words and their associated phones. The detailed pronunciation score assignments for the free-speaking scenario are provided in Appendix D.

---
[5] https://github.com/bicheng1225/HIPPO/tree/main

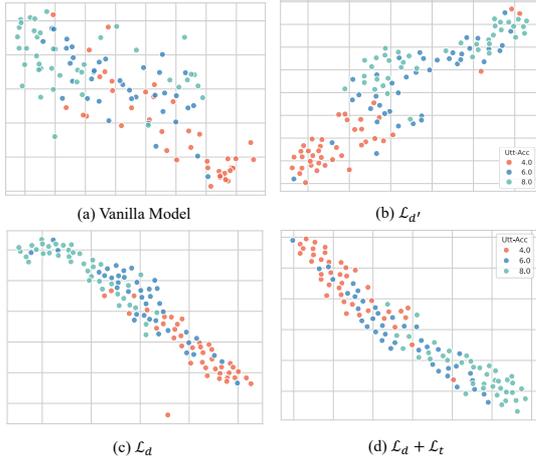

Figure 6: Visualization of utterance-level representations $Z$, where the orange, blue, and green points indicate accuracy scores of 4.0, 6.0, and 8.0, respectively. The plots display feature points for: (a) vanilla model, (b) vanilla model with a modified diversity term $\mathcal{L}_{d'}$ where the penalty is removed, (c) vanilla model with diversity term $\mathcal{L}_d$, and (d) vanilla model with CONO regularizer $\mathcal{L}_{CONO}$.

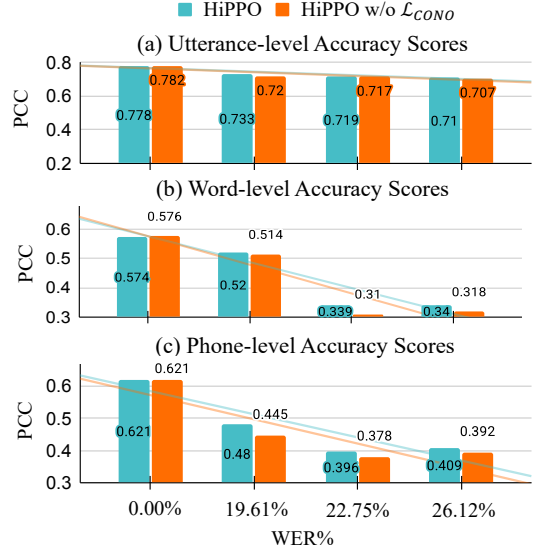

Figure 7: A comparison of PCC scores for pronunciation accuracy at the phone, word, and utterance levels between HiPPO and HiPPO w/o $\mathcal{L}_{Cono}$ under varying word error rate (WER) conditions. These WERs are calculated based on the reference text and different input transcriptions which are reference text and outputs of Whisper models (viz., large-v3, medium-en, small-en).

**Evaluation Metrics.** 1) Pearson correlation coefficient (PCC, ↑) measures the linear correlation between predicted and ground-truth scores for disparate pronunciation aspects. 2) Mean squared error (MSE, ↓) evaluates score discrepancy of the phone-level accuracy. The mean and standard deviation are reported for both metrics.

## 4 Experimental Results

**Assessments in the Free-speaking Scenarios.** At the outset, we compare our HiPPO with several current top-of-the-line APA models in the simulated free-speaking scenarios. From the results shown in Table 1, we make the following observations. 1) Our HiPPO achieves better PCC scores than all other competitive methods across different pronunciation aspects and linguistic granularities. 2) As to the ASR-free models, both VanillaSSL and Liu2023 are limited to utterance-level assessment, lacking finer-grained aspect scores at the phone or word level. Moreover, Liu2023 outperforms VanillaSSL in assessing the utterance-level fluency, where the gains stem from the integration of frame-level phonetic information via $k$-means clustering. Note also that effectively using phonetic information to boost assessment performance has been verified in prior work (Gong et al., 2022). Subsequently, compared to MultiPA, our method extracts pronunciation feature at the phone-level and then qualifies pronunciation aspects hierarchically across linguistic granularities, resulting in superior assessment performance. 3) In comparison among the variants of HiPPO, Parallel-CTC and Parallel-LLaMA outperform Hier-LLaMA in most assessment tasks. This observation suggests that, when pronunciation features are extracted from the transcripts containing ASR errors, the parallel design offers a more flexible and robust neural architecture for assessments in free-speaking scenarios compared to the hierarchical one. Notably, HiPPO stands out in assessment performance via the synergy of CONO regularizer and curriculum learning strategy.

| Models | Phone Scores | | Word Score (PCC) | | Utterance Score (PCC) | | | |
|---|---|---|---|---|---|---|---|---|
| | MSE↓ | PCC↑ | Accuracy↑ | Total↑ | Accuracy↑ | Fluency↑ | Prosody↑ | Total↑ |
| AzurePA | - | - | 0.623 | - | 0.700 | 0.715 | **0.842** | 0.782 |
| GOPT | 0.085 (±0.001) | 0.612 (±0.003) | 0.533 (±0.004) | 0.549 (±0.002) | 0.714 (±0.004) | 0.753 (±0.008) | 0.760 (±0.006) | 0.742 (±0.005) |
| 3M | **0.078** (±0.001) | 0.656 (±0.005) | 0.598 (±0.005) | 0.617 (±0.005) | 0.760 (±0.004) | 0.828 (±0.006) | 0.827 (±0.008) | 0.796 (±0.005) |
| HiPAMA | 0.084 (±0.001) | 0.616 (±0.004) | 0.575 (±0.004) | 0.591 (±0.004) | 0.730 (±0.002) | 0.749 (±0.001) | 0.751 (±0.002) | 0.754 (±0.002) |
| HierTFR | 0.081 (±0.000) | 0.644 (±0.000) | 0.622 (±0.002) | 0.634 (±0.002) | 0.735 (±0.008) | 0.801 (±0.004) | 0.795 (±0.002) | 0.764 (±0.002) |
| Parallel-TFR | 0.078 (±0.001) | 0.650 (±0.009) | 0.575 (±0.018) | 0.589 (±0.013) | 0.754 (±0.011) | 0.816 (±0.006) | 0.806 (±0.007) | 0.772 (±0.010) |
| Parallel-LLaMA | 0.074 (±0.002) | 0.658 (±0.007) | 0.598 (±0.012) | 0.610 (±0.009) | 0.774 (±0.009) | 0.837 (±0.006) | 0.829 (±0.004) | 0.796 (±0.009) |
| Hier-LLaMA | 0.082 (±0.002) | 0.656 (±0.006) | 0.622 (±0.006) | 0.634 (±0.008) | 0.789 (±0.006) | 0.844 (±0.003) | 0.832 (±0.003) | 0.811 (±0.005) |
| HiPPO* | 0.080 (±0.001) | **0.657** (±0.001) | **0.630** (±0.009) | **0.643** (±0.009) | **0.791** (±0.002) | **0.845** (±0.001) | 0.837 (±0.001) | **0.816** (±0.001) |

Table 2: The performance evaluations of our model and all compared methods on Speechocean762 test set in the read-aloud scenarios. HiPPO* refers to the model trained without curricular strategy and CONO regularizer.

**Ablation Studies in Free-speaking Scenarios.** As shown in the last two columns of Table 1, we ablate HiPPO with following settings: removing the CONO regularizer (w/o CONO) and substituting the curriculum learning strategy with training on a combined dataset of reading-aloud and free-speaking scenarios (w/o CL). From these ablation studies we can observe that both the CONO regularizer and the curriculum strategy are crucial to HiPPO. Removing either one of them leads to a decline in performance across several aspects and granularities. Second, the curriculum learning strategy makes a substantial contribution to the performance. Training HiPPO with the combined dataset, in contrast, results in lower performance across all assessment tasks.

**Qualitive Analysis on the CONO Regularizer in the Free-speaking Scenarios.** In Figure 6, we qualitatively examine the effectiveness of the additional training regularizer on the proposed hierarchical model. As depicted in Figure 6, the feature points in these subfigures display ordinal relationships, which are sorted by their utterance-level scores, with blue points being located between red points and green points. This result can be attributed to the aggregation of representations Z from the phone-level representations, which are highly correlated with the utterance-level accuracy score (Yan et al., 2024). By comparing Figures 6(b) with 6(c), it is evident that both diversity terms ($\mathcal{L}_{d'}$ and $\mathcal{L}_d$) can capture subtle differences between utterance-level scores, where feature points are clustered by their respective accuracy scores. The integration of ordinal penalty, as shown in Figure 6(c), further facilitates a clearer scattering of feature representations, with blue and green points more distinctly spread out. Finally, the impact of the tightness term $\mathcal{L}_t$ is verified in Figure 6(d), where the feature points exhibit tighter clustering in comparison with other subfigures.

**Effectiveness of CONO Regularizer across Different ASR Word Error Rate Settings.** Figure 7 examines the effectiveness of CONO regularizer $\mathcal{L}_{Cono}$ for the assessment accuracy at different granularities across various ASR word error rates (WERs), by comparing the HiPPO and its ablated version (HiPPO w/o CONO). Notably, in this set of experiments, our models were trained on the reference text and transcripts generated by whisper-large-v3 (achieving a WER of 19.6%) via proposed curricular learning strategy. First, with reference text as the input transcript, the assessment performance of both models seems comparable across granularities (phone, word, and utterance levels). Second, at the utterance-level assessment, the PCC scores of these two models appear relatively immune to WER degradation. A possible reason is that utilization of SSL-based features in utterance-level modeling, as the SSL models are often pre-trained on complex acoustic environments. Finally, the benefits of the CONO regularizer are more prominent at finer-grained linguistic levels. Specifically, the performance degrades substantially at the phone and word levels; however, the performance of HiPPO exhibits a more attenuated decline in comparison to other variants, which highlights the robustness of the proposed regularizer to ASR errors.

**Assessments in the Read-aloud Scenario.** In Table 2, the proposed HiPPO is evaluated in a read-aloud setting, where reference texts are employed in training and test. The main findings are presented as follows. 1) HiPPO markedly outperforms other methods in most pronunciation aspects. Notably, in contrast to prior studies, i.e., parallel models (GOPT and 3M) and hierarchical ones (HiPAMA and HierTFR), our model assesses pronunciation quality without explicit phone-level timestamps and achieves superior performance across various pronunciation aspects. 2) AzurePA stands out at the assessment of utterance-level prosody, whereas its performance on the other pronunciation aspects trails behind that of the other methods. These inferior results probably stem from that AzurePA is a commercial system that might has not been finetuned on Speechocean762. 3) As to the comparison between the variants of HiPPO (Parallel-LLaMA, Parallel-TFR, and Hier-LLaMA), Hier-LLaMA attains superior performance in most pronunciation aspects, particularly at the word and utterance levels, with a slight sacrifice in performance at the phone-level. These results are in line with the findings from previous studies (Do et al., 2023; Chao et al., 2023). By comparing HiPPO with Hier-LLaMA, we can verify that the proposed Conv-LLaMA block brings consistent improvements to pronunciation assessments.

## 5 Conclusion

In this paper, we have proposed a novel hierarchical pronunciation assessment model (dubbed HiPPO) for the spoken languages. To address arbitrarily long pronunciation feature sequences and capturing articulation traits across various linguistic granularities, we designed a Conv-LLaMA block for the proposed model. A contrastive ordinal regularizer is put forward to enhance robustness against ASR errors. Moreover, we explored a simple yet effective curriculum learning strategy for the spoken language assessment. Extensive experimental results validate the feasibility and effectiveness of the proposed methods, obtaining superior assessment performance compared to several state-of-the-art methods in both reading-aloud and stimulated free-speaking scenarios. In future work, we plan to explore more robust assessment models under various word error rate conditions for unscripted pronunciation assessments.

## 6 Limitations

Spoken language assessment gauges language competence across three sub-dimensions: pronunciation (fluency and delivery), language use (vocabulary and grammar), and topic development (content and discourse). In this paper, however, HiPPO focuses exclusively on pronunciation assessment within the broader context of spoken language evaluation. The following are several limitations of HiPPO in real-world applications:

**Transcriptions Containing ASR Errors.** Although speech foundation models have achieved near-human accuracy on public benchmark datasets, transcribing non-native English speech remains challenging. In our experiments, the word error rate (WER) for Speechocean762, transcribed using Whisper-large-v3, is 19.22% for the training set and 17.49% for the test set. Examining the performance of HiPPO through the lens of different WER conditions, we observed a significant degradation when ASR errors were severe, even with the proposed CONO regularizer.

**Lack of Accent Diversity.** The used dataset merely contains Mandarin L2 learners, hindering the generalizability of the proposed model and could be untenable when assessing the L2 learners with diverse accents.

**The Lack of Interpretability.** The model of the proposed method simply trains to mimic expert's annotations without resorting to manual assessment rubrics or other external knowledge, making it not straightforward to provide reasonable explanations for the assessment performance.

**Ethics Statement**

We hereby acknowledge that all of the co-authors of this work compile with the provided ACL Code of Ethics and honor the code of conduct. Our experimental corpus, Speechocean762, is widely used and publicly available. We think there are no potential risks for this work.

# References

Stefano Bannò and Marco Matassoni. 2022. Proficiency assessment of L2 spoken English using wav2vec 2.0. In Proceedings of the IEEE Spoke Language Technology Workshop (SLT), pages 1088–1095.

Yoshua Bengio, Jérôme Louradour, Ronan Collobert, and Jason Weston. 2009. Curriculum learning. In Proceedings of the International Conference on Machine Learning (ICML), pages 41–48.

Xinwei Cao, Zijian Fan, Torbjørn Svendsen, Giampiero Salvi. 2024. A framework for phoneme-level pronunciation assessment using CTC. In Proceedings of Interspeech (INTERSPEECH), pages 302–305.

Thibault Castells, Philippe Weinzaepfel, and Jerome Revaud. 2020. Superloss: A generic loss for robust curriculum learning. In Proceedings of Advances in Neural Information Processing Systems (NeurIPS).

Fu-An Chao, Tien-Hong Lo, Tzu-I. Wu, Yao-Ting Sung, Berlin Chen. 2022. 3M: An effective multi-view, multigranularity, and multi-aspect modeling approach to English pronunciation assessment. In Proceedings of the Asia-Pacific Signal and Information Processing Association Annual Summit and Conference (APSIPA ASC), pages 575–582.

Fu-An Chao, Tien-Hong Lo, Tzu-I. Wu, Yao-Ting Sung, Berlin Chen. 2023. A Hierarchical Context-aware Modeling Approach for Multi-aspect and Multigranular Pronunciation Assessment. In Proceedings of Interspeech (INTERSPEECH), pages 974–978.

Yu-Wen Chen, Zhou Yu, and Julia Hirschberg. 2024. MultiPA: A multi-task speech pronunciation assessment model for open response scenarios. In Proceedings of Interspeech (INTERSPEECH), pages 297–301.

Eduardo Coutinho, Florian Hönig, Yue Zhang, Simone Hantke, Anton Batliner, Elmar Nöth, and Björn Schuller. 2016. Assessing the Prosody of Non-Native Speakers of English: Measures and Feature Sets. In Proceedings of the Tenth International Conference on Language Resources and Evaluation (LREC), pages 1328–1332.

Huaijin Deng, Youchao Lin, Takehito Utsuro, Akio Kobayashi, Hiromitsu Nishizaki, and Junichi Hoshino. 2020. Automatic fluency evaluation of spontaneous speech using disfluency based features. In Proceedings of the IEEE International Conference on Acoustics, Speech and Signal Processing (ICASSP), pages 9239-9243,

Heejin Do, Yunsu Kim, and Gary Geunbae Lee. 2023. Hierarchical pronunciation assessment with multi-aspect attention. In Proceedings of the IEEE International Conference on Acoustics, Speech and Signal Processing (ICASSP), pages 1–5.

Keelan Evanini, Maurice Cogan Hauck, and Kenji Hakuta. 2017. Approaches to automated scoring of speaking for K–12 English language proficiency assessments. ETS Research Report Series, pages 1–11.

Nancy F. Chen and Haizhou Li. 2016. Computer-assisted pronunciation training: From pronunciation scoring towards spoken language learning. In Proceedings of the Asia-Pacific Signal and Information Processing Association Annual Summit and Conference (APSIPA ASC), pages 1–7.

Luciana Ferrer, Harry Bratt, Colleen Richey, Horacio Franco, Victor Abrash, Kristin Precoda. 2015. Classification of lexical stress using spectral and prosodic features for computer-assisted language learning systems. Speech Communication, volume 69, pages 31–45.

Horacio Franco, Harry Bratt, Romain Rossier, Venkata Rao Gadde, Elizabeth Shriberg, Victor Abrash, and Kristin Precoda. 2010. EduSpeak: A speech recognition and pronunciation scoring toolkit for computer-aided language learning applications. Language Testing, volume 27, pages 401–418.

Yuan Gong, Ziyi Chen, Iek-Heng Chu, Peng Chang, and James Glass. 2022. Transformer-based multi-aspect multigranularity non-native English speaker pronunciation assessment. In Proceedings of the IEEE International Conference on Acoustics, Speech and Signal Processing (ICASSP), pages 7262–7266.

Wenping Hu, Yao Qian, Frank K. Soong, and Yong Wang. 2015. Improved mispronunciation detection with deep neural network trained acoustic models and transfer learning based logistic regression classifiers. Speech Communication, volume 67, pages 154–166.

Yassine Kheir, Ahmed Ali, and Shammur Chowdhury. 2023. Automatic Pronunciation Assessment–A Review. In Findings of the Association for Computational Linguistics: EMNLP, pages 8304–8324.

Eesung Kim, Jae-Jin Jeon, Hyeji Seo, Hoon Kim. 2022. Automatic pronunciation assessment using self-supervised speech representation learning. In Proceedings of Interspeech (INTERSPEECH), pages 1411–1415.

Yaman Kumar Singla, Avyakt Gupta, Shaurya Bagga, Changyou Chen, Balaji Krishnamurthy, Rajiv Ratn Shah. 2021. Speaker-conditioned hierarchical modelling for automated speech scoring. In Proceedings of the ACM International Conference on Information & Knowledge Management (CIKM), pages 1681–1691.

Binghuai Lin and Liyuan Wang. 2021. Deep feature transfer learning for automatic pronunciation assessment. In Proceedings of Interspeech (INTERSPEECH), pages 4438–4442.

Wei Liu, Kaiqi Fu, Xiaohai Tian, Shuju Shi, Wei Li, Zejun Ma, and Tan Lee. 2023. An ASR-free fluency scoring approach with self-supervised learning. In Proceedings of the IEEE International Conference on Acoustics, Speech and Signal Processing (ICASSP), pages 1–5.

Tien-Hong Lo, Fu-An Chao, Tzu-I Wu, Yao-Ting Sung, and Berlin Chen. 2024. An effective automated speaking assessment approach to mitigating data scarcity and imbalanced distribution. In Findings of the Association for Computational Linguistics: NAACL, pages 1352–1362.

Pamela M Rogerson-Revell. 2021. Computer-assisted pronunciation training (CAPT): Current issues and future directions. RELC Journal, volume 52, pages 189–205.

John M. Norris and Larry Davis. 2025. Assessing second language speaking at ETS: Introduction. In: Challenges and Innovations in Speaking Assessment. Routledge, pages 1–18.

Silke M. Witt and S. J. Young. 2000. Phone-level pronunciation scoring and assessment for interactive language learning. Speech Communication, volume 30, pages 95–108.

Pieter Mülller, Febe de Wet, Christa van der Walt, and Thomas Niesler. 2009. Automatically assessing the oral proficiency of proficient L2 speakers. In Workshop on Speech and Language Technology in Education (SLaTE), pages 29–32.

Hao-Chen Pei, Hao Fang, Xin Luo, Xin-Shun Xu. 2024. Gradformer: A framework for multi-aspect multi-granularity pronunciation assessment. IEEE/ACM Transactions on Audio, Speech, and Language Processing, volume 32, pages 554–563.

Yifan Peng, Siddharth Dalmia, Ian Lane, and Shinji Watanabe. 2022. Branchformer: Parallel mlp-attention architectures to capture local and global context for speech recognition and understanding. In International Conference on Machine Learning (PMLR), pages 17627–17643

Yang Shen, Ayano Yasukagawa, Daisuke Saito, Nobuaki Minematsu, and Kazuya Saito. 2021. Optimized prediction of fluency of L2 English based on interpretable network using quantity of phonation and quality of pronunciation. In Proceedings of IEEE Spoken Language Technology Workshop (SLT), pages 698–704.

Jianlin Su, Murtadha H. M. Ahmed, Yu Lu, Shengfeng Pan, Wen Bo, and Yunfeng Liu. 2024. Roformer: Enhanced Transformer with rotary position embedding. Neurocomputing, volume 568.

Hugo Touvron, Louis Martin, Kevin Stone, Peter Albert, Amjad Almahairi, et al. 2023. Llama 2: Open foundation and fine-tuned chat models. arXiv preprint arXiv:2307.09288.

Nidhi Vakil and Hadi Amiri. 2023. Curriculum Learning for Graph Neural Networks: A Multiview Competence-based Approach. In Proceedings of the Annual Meeting of the Association for Computational Linguistics (ACL), pages 7036–7051.

Alistair Van Moere and Ryan Downey. 2016. Technology and artificial intelligence in language assessment. Handbook of second language assessment, pages 341–358.

Yihao Wang, Zhongdi Wu, Joseph Nese, Akihito Kamata, Vedant Nilabh, Eric C. Larson. 2025. A unified model for oral reading fluency and student prosody. In Proceedings of the IEEE International Conference on Acoustics, Speech and Signal Processing (ICASSP), pages 1–5.

Ke Wang, Lei He, Kun Liu, Yan Deng, Wenning Wei, Sheng Zhao. 2025b. Exploring the potential of large multimodal models as effective alternatives for Pronunciation assessment. in arXiv preprint arXiv:2503.11229.

Benjamin Warner, Antoine Chaffin, Benjamin Clavié, Orion Weller, Oskar Hallström, et al. 2024. Smarter, better, faster, longer: A modern bidirectional encoder for fast, memory efficient, and long context finetuning and inference. arXiv preprint arXiv:2412.13663.

Bi-Cheng Yan and Berlin Chen. 2024. An effective hierarchical graph attention network modeling approach for pronunciation assessment. IEEE/ACM Transactions on Audio, Speech, and Language Processing, volume 32, pages 3974–3985.

Bi-Cheng Yan, Jiun-Ting Li, Yi-Cheng Wang, Hsin-Wei Wang, Tien-Hong Lo, Yung-Chang Hsu, Wei-Cheng Chao, and Berlin Chen. 2024. An Effective Pronunciation Assessment Approach Leveraging Hierarchical Transformers and Pre-training Strategies. In Proceedings of the Annual Meeting of the Association for Computational Linguistics (ACL), pages 1737–1747.

Klaus Zechner, and Keelan Evanini. 2019. Automated speaking assessment: Using language technologies to score spontaneous speech. Routledge.

Biao Zhang and Rico Sennrich. 2019. Root mean square layer normalization. In Proceedings of Advances in Neural Information Processing Systems (NeurIPS), volume 32.

Junbo Zhang, Zhiwen Zhang, Yongqing Wang, Zhiyong Yan, Qiong Song, Yukai Huang, Ke Li, Daniel Povey, and Yujun Wang. 2021. Speechocean762: An open-source non-native English speech corpus for pronunciation assessment. In Proceedings of Interspeech (INTERSPEECH), pages 3710 –3714.

Jian Zhu, Cong Zhang, and David Jurgens. 2022. Phone-to-audio alignment without text: A semi-supervised approach. In Proceedings of the IEEE International Conference on Acoustics, Speech and Signal Processing (ICASSP), pages 8167–8171.

## A  Related Work

Automatic Pronunciation Assessment (APA) quantifies L2 learners' pronunciation proficiency in a target language, offering either analytic scores (continuous numerical values for specific aspects) or an overall score (discrete categorical values for speaking competence). We categorize the related APA works into the following two groups for discussion, differentiated by their reliance on reference text.

**Scripted-speech Assessment.** The developments of scripted-speech assessment are typically designed in read-aloud learning scenarios, where an L2 learner is provided with a reference text and instructed to pronounce it verbatim. Early efforts in scripted speech assessment predominantly focused on single-aspect assessment, which predicted proficiency scores at specific linguistic levels with various sets of hand-crafted features by separate scoring modules, such as phone-level accuracy (Witt and Young, 2000), word-level stress (Ferrer et. al., 2015), and utterance-level fluency (Coutinho et. al., 2016). Furthermore, the commonly used hand-crafted features were derived from the reference text in conjunction with the learner's speech via an ASR model (hybrid DNN-HMM system), where the extracted pronunciation features included acoustic features, confidence scores of recognized linguistic units, time-alignment information, and statistical measures, but were not limited to these (Mülller et al., 2009; Franco et al., 2010). To provide comprehensive pronunciation feedback for language learners, a flurry of recent work has advocated multi-aspect and multi-granular pronunciation assessment, which evaluates pronunciation proficiency across multiple linguistic levels (viz. phoneme, word, and utterance), with diverse aspects (e.g., accuracy, fluency, and completeness) with a unified model. Drawing on this research trend, various neural models capitalizing on hand-crafted features and/or self-supervised features have been extensively investigated in existing literature, including parallel (Gong et al., 2022; Chao et al., 2022), hierarchical (Do et al., 2023, Yan et al., 2024), and linguistic-decoupled structures (Pei et al., 2024).

**Unscripted-speech Assessment.** Unscripted-speech assessment is an emerging research field and has gained increasing attention in recent years, as it attempts to qualify learners' speaking abilities in real-world communication. The corresponding developments target free-speaking scenarios, in which an L2 learner receives a reference text (with short questions) and is expected to respond or share opinions grounded in their personal experiences. Based on the free-form and spontaneous speech inputs, the assessment models then qualify oral skills and provide instructive feedback at various linguistic levels. As one of the initial attempts, Liu et al. (2023) proposed an ASR-free method which leveraged a pre-trained self-supervised learning (SSL) model (viz., wav2vec2.0) to estimate fluency scores for L2 learners without resorting to the reference texts (or ASR transcriptions). Their method extracted frame-level acoustic features with a self-supervised learning (SSL) model and generated phonology features by assigning proximity phone labels (cluster index) to each frame via K-means clustering. To assess word-level speaking skills, a pioneering effort, Chen et al. (2024) proposed MultiPA which extracted pronunciation features at word-level with two speech recognizers and employed a bottom-up neural structure to examine the learner's pronunciation skills at both word and utterance levels. Specifically, one recognizer employs a high-performing ASR model (whisper-large-v3) to approximate the ground-truth word sequence, while the other utilizes an ASR model trained with native speaker speech (whisper-medium-en) to emulate how a native speaker would process the learner's speech. Compared to previous works, our model assesses oral skills from phone-level to utterance-level by working in tandem with a speech recognizer and a G2P converter. Furthermore, to mitigate the detrimental effects of ASR errors, we proposed the extract score-discriminative features by leveraging the contrastive ordinal regularizer.

## B Implementation Details

This section illustrates the implementation details of our experiments, and we plan to make our source code and datasets publicly accessible after the reviewing process.

**Training Hyperparameters.** Our implementation follows previous studies (Gong et al., 2022; Chao et al., 2022), employing the Adam optimizer, with a learning rate of 0.001, and a batch size of 25. To stabilize the training process, the aspect scores at both the utterance and word levels are normalized to match the scale of the phone-level score, ranging from 0 to 2. We conducted 5 independent trials, with each trial running for 100 epochs and using a different random seed to reduce the impact of randomness. The evaluation metrics are reported as the average of the best-performing epochs across these trials, selected based on the minimum phone-level MSE values.

**Model Configurations.** In the Conv-LLaMA block, the multi-head self-attention (MHSA) module is configured with 1 head and 24 hidden units for different granularities ($d_\text{p} = d_\text{w} = d_\text{u} = 24$). The attention pooling mechanisms at the word and utterance levels share the same configuration, which use a single-layer multi-head attention mechanism with 3 heads and 24 hidden units. Furthermore, in each modeling stage, the regressors for various pronunciation aspects are implemented as feed-forward networks, each consisting of two linear transforms with a non-linear activation in between, and the second transform of each projects the hidden dimension to a single scalar output.

## C Comparative Methods

We compare HiPPO with several top-of-the-line methods in both simulated free-speaking and read-aloud scenarios.

**Comparative Models for Simulated Free-speaking Assessment.** First, for the free-speaking scenario, we compare three categories of methods. 1) ASR-free methods: VanillaSSL (Chen et al., 2024) qualifies utterance-level pronunciation aspects by fine-tuning a pre-trained self-supervised learning (SSL) model (viz., wav2vec2.0); similarly, Liu et al. (2023), based on a SSL-based acoustic model, first extracts frame-level SSL features, subsequently assigning phonetic information to each frame with k-means clustering and then evaluating utterance-level pronunciation aspects via a simple mean pooling mechanism. 2) ASR-based method: MultiPA (Chen et al., 2024) extracts pronunciation features based on two speech recognizers and constructs an assessment model to qualify oral skills at word and utterance levels. 3) Variants of HiPPO: Parallel-CTC, Parallel-LLaMA, and Hier-LLaMA adopt the same inputs as HiPPO (i.e., $X^\text{p}$ and $\mathbf{x}^\text{ssl}$), while exploring different neural architectures. Parallel-CTC and Parallel-LLaMA adopt a parallel architecture, with the former using Transformer blocks and the latter stacking of LLaMA blocks. Hier-LLaMA replaces the Conv-LLaMA blocks of HiPPO with standard LLaMA blocks.

**Comparative Models for Read-aloud Assessment.** For read-aloud scenario, we first report the performance of Azure Pronunciation Assessment (AzurePA) service (Wang et al., 2025b), followed by a comparison with several APA models. 1) Parallel neural structure: GOPT (Gong et al., 2022) adopts pronunciation features derived from phone-level timestamps and models the pronunciation aspects with Transformer blocks; 3M (Chao et al., 2022) extends GOPT by incorporating acoustic features, i.e., phone duration statistics and SSL-based features, and phonology features, i.e., vowel and consonant embeddings. 2) Hierarchical neural structure: HiPAMA (Do et al., 2023) is a language hierarchy-aware APA model equipped with the trait attention mechanisms; HierTFR (Yan et al., 2024) stacked a hierarchical neural structure via Transformer blocks and proposed mask prediction to strengthen the relationships across granularities for model initialization.

## D Score Assignments for Speechocean762 Corpus in the Simulated Free-speaking Scenario

Speechocean762 was curated in a read-aloud learning scenario, where human annotators provided pronunciation scores at the utterance, word, and phone levels. These scores were assigned based on the reference text (for scoring utterance-level and word units) and the corresponding canonical phone sequence (for scoring phone units). To reorganize

Speechocean762 in the simulated free-speaking scenario, we first use a speech foundation model (whisper-large-v3) to transcribe the learners' speech and convert the transcriptions into the corresponding phone sequences via a G2P converter (g2pE). Next, for each recording, we align the ASR transcription to reference text and the converted phone sequence to canonical phone sequence, respectively. Based on the resulting alignments, we first assigned pronunciation scores from human annotators to correctly recognized segments (i.e., including phone and word units). For subsequent score assignments, we handled ASR errors at both levels (viz., word and phone levels) as follows:

- Deletion Errors: Ignored due to there are no corresponding segments in the transcribed words (or converted phones).
- Substitution Errors: Assigned scores based on aligned segments, as most substitution error cases reflect subtle acoustic differences.
- Insertion Errors: Assigned a score of zero.

Figure 8 presents the alignment process for a sample recording. Note that insertion errors are retained, owing to the maintenance of phone-to-word mappings for developing hierarchical neural structure. Figure 8(c) highlights how the score assignment process maintains phone-to-word relationships for converted phones (i.e., the mapping of phone segments $p_4$ and $p_7$ to word G).

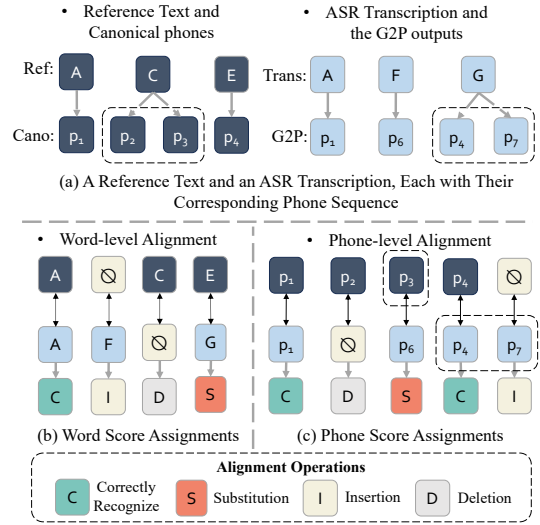

Figure 8: Illustration of score assignments for Speechocean762 in simulated free-speaking scenarios. For a sample recording, we demonstrate the assignment process: (a) shows the reference text and an ASR transcription, along with their respective canonical and G2P converted phone sequences; (b) alignment of ASR transcription to reference text for human scoring; (c) alignment of G2P outputs to canonical phones for pronunciation scoring.